\documentstyle[preprint,aps,epsf]{revtex}
\begin{document}
\preprint{SSF96-02-01}
\tighten
\title{The exclusive (e,e$'$p) reaction at high missing momenta}
\author{V. Van der Sluys, J. Ryckebusch\   and M. Waroquier\\ 
Institute for Theoretical
Physics and Institute for Nuclear Physics\protect\\
 Proeftuinstraat 86 \protect\\
B-9000 Gent, Belgium}

\date{\today}

\maketitle

\begin{abstract}
\noindent 
The reduced (e,e$'$p) cross section  is
calculated for kinematics that probe high missing momenta. 
The final-state
interaction is handled within a non-relativistic many-body framework. 
One- and two-body nuclear currents are included.
Electron distortion effects are treated in an exact distorted wave
calculation.
It is shown that 
at high missing momenta the calculated (e,e$'$p) cross sections exhibit a
pronounced sensitivity to ground-state correlations of the RPA type
and two-body currents.  The role of these mechanisms  is found to be relatively
small at low missing momenta.
\end{abstract}


\section{Introduction}

For long the exclusive (e,e$'$p) reaction for quasielastic (QE)
kinematics
 has been considered as the ideal
tool to study
the single-particle properties of the nucleus
\cite{fru84,bof93,kel95}.
One of the most remarkable
conclusions drawn from the QE (e,e$'$p) data was related to  the
low extracted spectroscopic factors which
could not be explained within the
independent particle model (IPM). 
Long- and short-range correlations beyond the IPM are found to
redistribute the hole strength over a range of excitation 
energies and to modify the nucleon momentum distribution with 
respect to the IPM prediction \cite{ben93,mah91}. 

Recently,  renewed interest 
has been observed for the exclusive (e,e$'$p) reaction.
With the new generation 
of electron facilities the high-momentum components of the cross
section have come into reach of experimental exploration.
The first results of this type of  experiments are available for
the $^{16}$O(e,e$'$p) \cite{blo95} and $^{208}$Pb(e,e$'$p)
\cite{bob94}
 reactions.
Data have been collected for proton knockout from the
single-particle orbits near the Fermi level.
The 
(e,e$'$p) reaction at high missing momenta is expected to provide
 direct information on the
nucleon momentum distribution at high momenta.  As such the reaction
could reveal information
about the interaction of the nucleons at
short internucleon distances.
One should, however, not forget that :
\begin{itemize}
\item{Various studies  
\cite{mut94,pol95,mut95,ben94,vne95,ant95,gai95} have shown
that the single-particle
spectral function gets strongly modified by short-range correlations
(SRC) at
large excitation energies with only marginal effects at low energies.
Accordingly, it is to be expected that short-range 
correlations hardly
affect the exclusive (e,e$'$p) knockout from the valence shells. 
}
\item{The (e,e$'$p) cross section is only directly related
to the single-particle spectral function in the Plane Wave Impulse
Approximation (PWIA). 
The PWIA approach encompasses several assumptions:
(i) The
nuclear current operator 
is constrained to be a one-body operator. (ii) No
final-state interaction (FSI) between the ejected nucleon and 
the residual nucleus
is taken into account.
(iii) Electron distortion effects are neglected. This assumption
becomes questionable for heavy target nuclei.}
\end{itemize}
It is clear that in order to extract reliable information on the
nucleon spectral
function from the (e,e$'$p) reaction,
the reaction mechanism has to be well understood.
Our theoretical
framework for the (e,e$'$p) reaction goes 
beyond the PWIA and addresses the final-state interaction
of the ejected nucleon with the residual core within a 
consistent many-body framework.
We include one- and two-body photoabsorption mechanisms.
On top of that the effect of Coulomb distortion on the electron wave
functions is included exactly.

The outline of this paper is as follows. In section II
we briefly discuss the adopted (e,e$'$p) formalism. Sections III
and IV deal with the results for the $^{16}$O(e,e$'$p) and
$^{208}$Pb(e,e$'$p) reactions. Finally, some conclusions are drawn in
section V.

\section{Formalism}

In this paper we describe the scattering process of an electron with energy
$\epsilon$ from a target nucleus, transferring an energy
$\omega$ and momentum $\vec{q}$ to the nuclear system. The energy
transfer $\omega$ is sufficient to eject a proton with momentum
$\vec{p}_p$ and spin projection $m_{s_p}$ out of the target nucleus.

In order to evaluate the (e,e$'$p) cross section one has to calculate
the nuclear response to the nuclear charge-current operator. In the
one-photon exchange approximation and 
for constant $\vec{q}$-$\omega$ kinematics, the unpolarized  
cross section
can be written in terms of a longitudinal, transverse,
longitudinal-transverse and transverse-transverse part:
\begin{eqnarray}
\frac{{\rm d}^6\sigma}{{\rm d}\epsilon'{\rm d}\Omega_e {\rm
d}\Omega_p {\rm d}T_p} &=& \sigma_L 
+ \sigma_T + \sigma_{LT} + \sigma_{TT} \;.
\end{eqnarray}
Each of these terms is related to the so-called nuclear structure functions
which in turn can be written as a specific combination of
matrixelements between the initial and residual nuclear system
 of the
charge-current four-vector $J_{\mu}(q)$ \cite{ras89}, i.e.,
\begin{eqnarray}
\label{matr}
<J_{R}M_{R};\vec{p}_{p},1/2m_{{s}_{p}}\mid 
J_{\mu}(q)\mid J_{i}M_{i}>
\end{eqnarray}
where  $\mid \! J_{i}M_{i}\! >$ and $\mid \! J_{R}M_{R}\! >$ describe
the target and residual nucleus.

At this point the nuclear structure model and the
photoabsorption mechanism need to be settled.
The initial nuclear state and the final-state interaction of 
the ejected proton with the residual
nucleus is treated self-consistently in the many-body formalism as outlined in
ref.~\cite{ryc88}. 
The final nuclear state is evaluated through a phase shift analysis
based on a partial-wave expansion.
The bound and continuum single-particle states are
determined with the same potential, i.e., a Hartree-Fock (HF)
 potential generated with
an effective interaction of the Skyrme type (SkE2) \cite{war87}.
We go beyond the independent particle model and 
incorporate long-range correlations not implemented in the
HF approach. 
This is achieved in the Random Phase Approximation (RPA).
The RPA states can be seen as a linear combination of
 one-particle one-hole
and one-hole one-particle excited states out of a correlated ground
state. The ground state of the target nucleus implicitly 
includes the long-range
two-particle two-hole correlations induced by 
the residual interaction. 

For the photoabsorption mechanism, we assume that the virtual photon 
is absorbed on one or two nucleons in the
nucleus.
This means that the nuclear current has a one- and two-body part:
$J_{\mu}(q)=J_{\mu}^{(1)}(q)+J_{\mu}^{(2)}(q)$.
The transverse 
nucleonic one-body current consists of the well-known convection and
magnetization current. The two-body current is taken from a
non-relativistic reduction of the lowest order Feynman diagrams with
one exchanged pion and intermediate $\Delta_{33}$-excitation. 
This procedure
gives rise to the well-known seagull terms, the pion-in-flight term
and terms with a $\Delta_{33}$-excitation in the intermediate state
\cite{ris89,vsl94}. In
this non-relativistic approach the nuclear charge operator is not
affected by two-body contributions.

For heavy target nuclei, the model has to account for
 electron distortion effects.
 The distortions of the initial and final electron
due to the static Coulomb field generated by the protons in the
nucleus, is treated exactly in a Coulomb Distorted Wave Born
Approximation (CDWBA) calculation \cite{ube71}. 
Details regarding the adopted numerical
procedure will be reported elsewhere \cite{vsl95}. 
When it comes to the treatment of electron distortions,
our approach is very similar in nature to the one
reported earlier by the Ohio and Madrid group \cite{jin1,jin2,udias2}.

As is commonly done in the analysis of the quasielastic 
(e,e$'$p) reaction,
the calculations and the data are plotted as a reduced 
cross section
which is derived from  the cross
section in the following way:
\begin{eqnarray}
\label{extheo}
\rho_m(p_m, E_x)=
\frac{1}{p_p E_p \sigma_{ep} }
\left(\frac{{\rm d}^6\sigma}{{\rm d}\epsilon'{\rm d}
\Omega_e {\rm d}\Omega_p {\rm d} T_p}\right) \;.
\end{eqnarray} 
Throughout this paper, we use the so-called CC1 
off-shell electron-nucleon cross
section $\sigma_{ep}=\sigma^{cc1}$ \cite{for83}.
The missing momentum is defined according to $\vec{p}_{m}=\vec{p}_p -
\vec{q}$. 
The sign convention for the 
missing momentum $p_m$ is chosen such that $p_m$ is 
negative for a proton ejected
in the half-plane determined by
 the initial electron momentum and bordered by the
momentum transfer. For proton ejection in the other half-plane,
 the missing momentum is
considered positive.
The excitation energy of the residual nucleus is denoted by $E_x$.

It should be stressed that only in the case that
 the FSI, two-body currents and
electron distortion can be neglected, the reduced cross section
coincides with the single-particle spectral function and can be
directly related to
the microscopic predictions for this quantity. In a similar way,
the missing
momentum $p_m$  can only be interpreted as  
the momentum of the nucleon before it was struck
by the virtual photon in the PWIA.
This means
that a careful examination of the various higher-order effects
needs to be performed before one can associate the high $p_m$ data
with the high-momentum components of the spectral function.
In this paper, the outlined formalism is applied to the exclusive
$^{16}$O(e,e$'$p) and $^{208}$Pb(e,e$'$p) reactions.

\section{Results for $^{16}$O(e,e$'$p)}

The $^{16}$O(e,e$'$p) reaction has been studied for different electron 
kinematics. 
As outlined in the previous section, 
our model implements two types of
nucleon-nucleon correlations beyond the IPM: long-range RPA
correlations, meson-exchange and isobaric contributions.
By varying the momentum transfer $q$ for a fixed
$\epsilon$ and $\omega$, we can investigate the
$q$-dependence of these additional components in the model.

\begin{figure}
\centering
\epsfxsize=8.cm
\epsffile{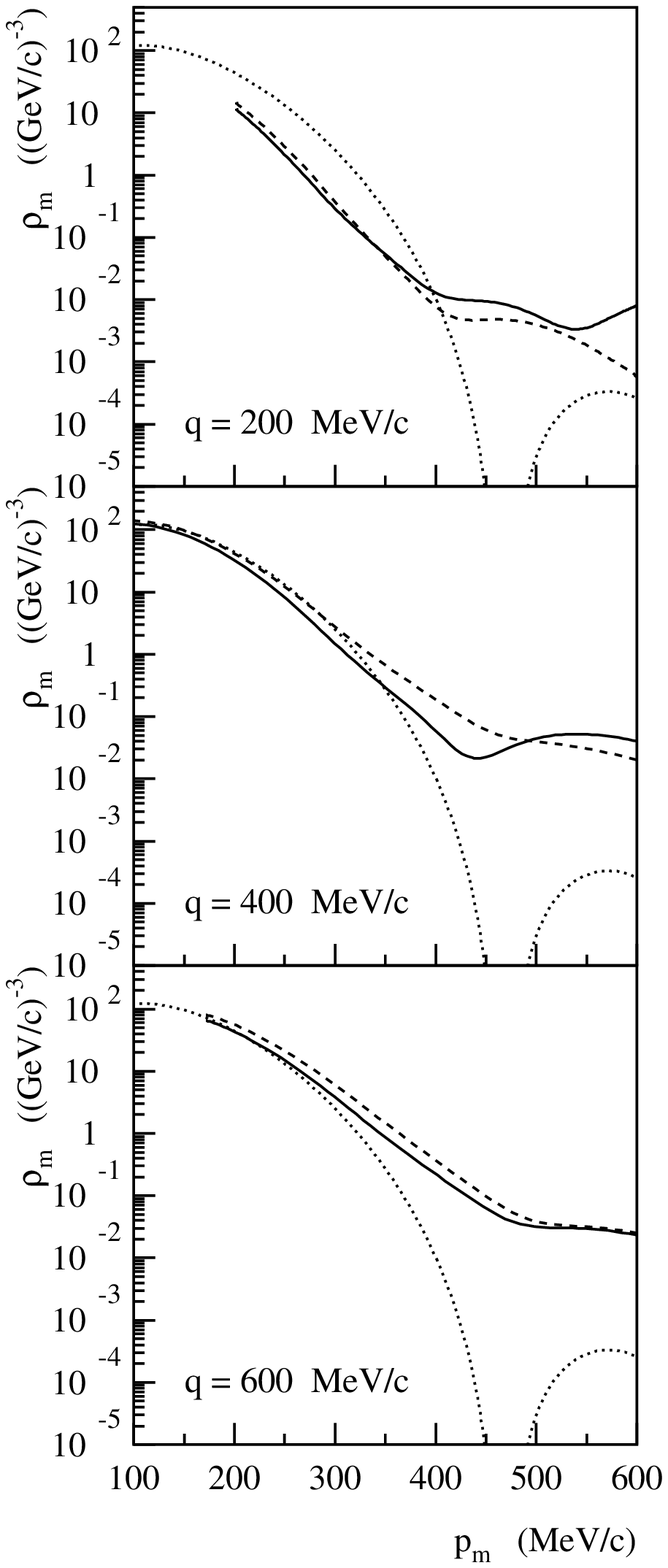}
\caption{The reduced cross section for proton knockout from the
$1p3/2$ shell in $^{16}$O for $\epsilon=500$ MeV, $\omega=100$ MeV
and different values of the momentum transfer. 
The dotted line represents the IPM momentum distribution, the dashed
line depicts 
the reduced cross section in the HF framework including only
one-body components in the nuclear current. For the solid line
long-range RPA correlations are accounted for.
The curves are not multiplied with a spectroscopic factor.  }
\label{fig1}
\end{figure}

In Fig.~\ref{fig1} we plot the reduced cross section for
electro-induced one-proton
knockout from the $1p3/2$ shell in $^{16}$O. 
Comparing the HF reduced cross section with the 
IPM momentum distribution,
a striking feature is that the final-state interaction of
the ejected proton with the residual nucleus generates 
a lot of strength at high $p_m$. Further, 
the HF reduced cross section has
a smoother $p_m$ dependence compared with the IPM
momentum distribution.
The long-range RPA correlations 
considerably modify the reduced cross
section at high $p_m$ and leave the low-momentum components of the
reduced cross section almost unaffected.
From Fig.~\ref{fig1} it is also clear that 
 for high-momentum transfer q, RPA-type of correlations
have a mariginal effect on the reduced cross section 
for the complete missing momentum range.

\begin{figure}
\centering
\epsfxsize=8.cm
\epsffile{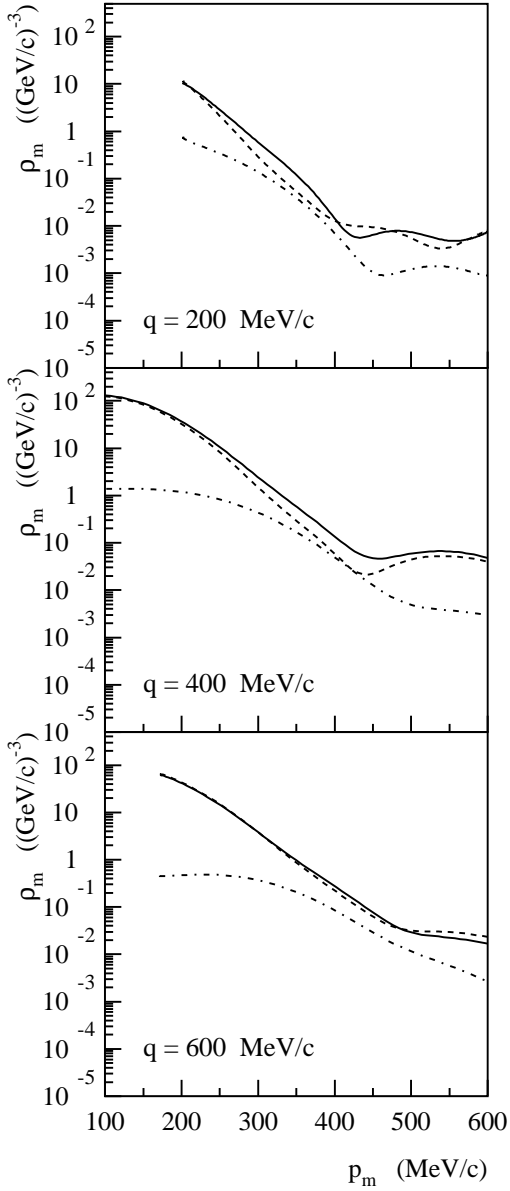}
\caption{The reduced cross section for proton knockout from the
$1p3/2$ shell in $^{16}$O for $\epsilon=500$ MeV, $\omega=100$ MeV
and different values of the momentum transfer. 
The dashed and dashed-dotted line give the
one-body respectively two-body (mesonic) contribution to the reduced cross
section.
The solid line is the coherent sum of these two contributions.
The curves are not multiplied with a spectroscopic factor.}
\label{fig2}
\end{figure}

The contribution of the two-body nuclear currents in the model is
investigated in  Fig.~\ref{fig2} and is found to exhibit a very
 similar behaviour for the
three considered kinematics. 

\begin{figure}
\centering
\epsfxsize=8.cm
\epsffile{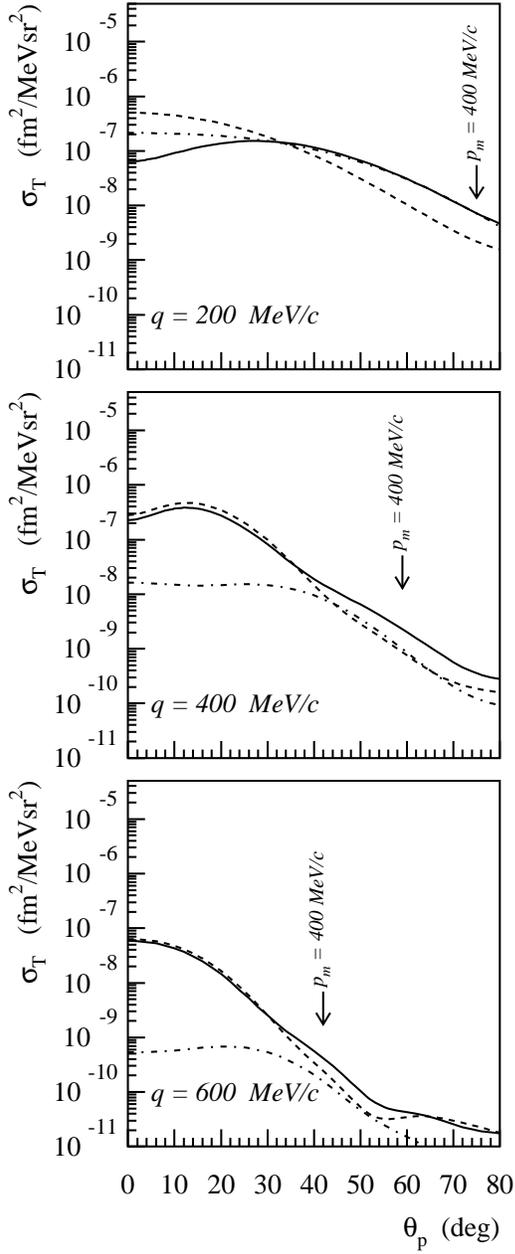}
\caption{The transverse cross section $\sigma_T$ for the
$^{16}$O(e,e$'$p) reaction for the same kinematics as in
Fig.~\protect\ref{fig2}\protect. 
We adopt the line conventions as defined in
Fig.~\protect\ref{fig2}\protect. 
The
curves are not multiplied with a spectroscopic factor.}
\label{fig3}
\end{figure}

In order to explain this behaviour 
we  firstly examine the transverse part $\sigma_T$ of the
cross section  in more detail as this quantity
is expected to show the largest sensitivity to the two-body currents. 
Fig.~\ref{fig3} clearly demonstrates that the contribution from two-body
nuclear currents becomes less important with increasing momentum
transfer. 
However, the high-momentum components of the reduced cross section
are probed in the tail of the transverse cross section where the
one-body current contribution becomes rather small. 
It
can be easily shown that for a fixed value of   the missing
momentum $p_m$ the  
proton scattering angle decreases with increasing $q$ 
(marked with arrows in Fig.~\ref{fig3}). 
Furthermore, in this scattering region, 
the two-body contribution to the cross section decreases with
increasing scattering angle. 
This feature can be easily explained combining that for higher
scattering angles (higher $p_m$) mainly high-momentum mesons are
exchanged between the nucleons and the fact that the meson propagator
decreases with increasing meson momentum.
The previous considerations allow 
us to conclude that
 even small contributions from 
mesonic currents come into
play when studying the high-momentum side of
the (e,e$'$p) cross section.

Secondly, the kinematics considered  in Figs.~\ref{fig1} and \ref{fig2}
 are  such that the energy transfer
and the energy of the incoming electron are fixed. This means
that the higher the momentum transfer, the more important the
transverse part of the cross section will be. As mesonic currents only
contribute to the transverse components of the cross section and do
not affect the longitudinal part, we can conclude that mesonic currents
gain in relative importance with increasing $q$.

Summarizing, we conclude that the considered $^{16}$O(e,e$'$p)
cross section is strongly affected by FSI and mesonic currents even for
relatively large $q$-values. 
Accordingly, this could seriously hamper the extraction of 
any information on the
single-particle spectral function.
A naive suggestion would be 
to study the high-momentum side of the cross section for
higher energy transfer since the FSI 
vanishes for $p_p \rightarrow
\infty$.
On the contrary, for these higher energy transfers it is to be expected
that isobaric currents gain in importance.

\section{Results for $^{208}$Pb(e,e$'$p)}

\begin{figure}
\centering
\epsfxsize=9.5cm
\epsffile{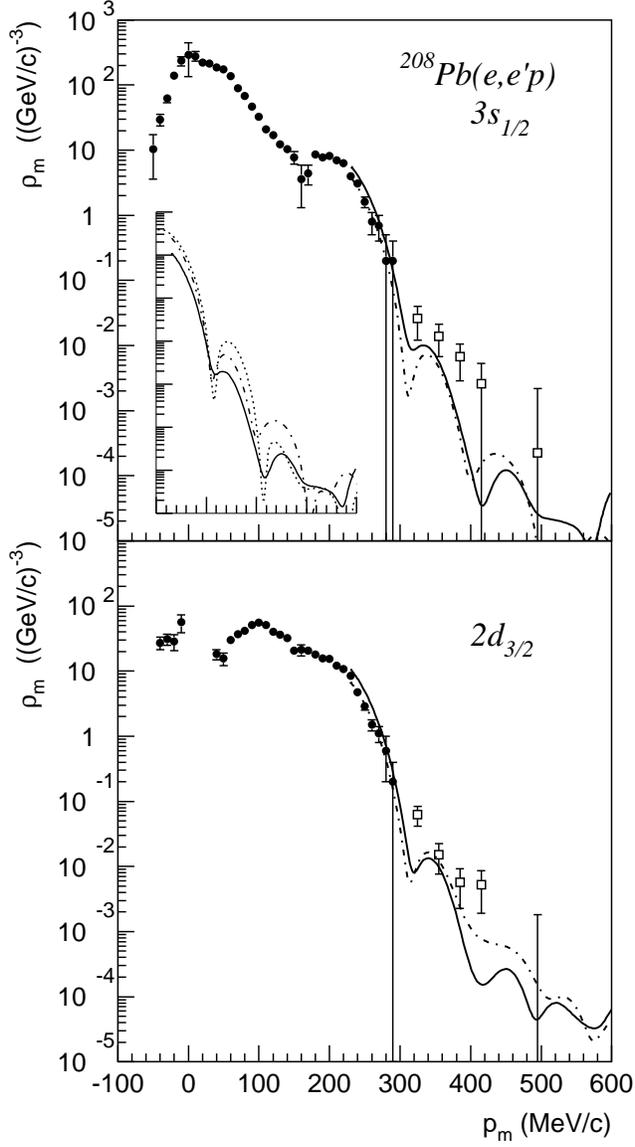}
\caption{The $^{208}$Pb(e,e$'$p) reduced cross sections for the
kinematics of the NIKHEF experiments~\protect\cite{bob94}\protect.
For the
dashed-dotted line the FSI is treated in the HF approach whereas for the
solid line the optical potential model is adopted. The optical
potential calculations are multiplied with the spectroscopic factors
($S(3s1/2)=0.51$, $S(2d3/2)=0.54$). The HF curves 
are multiplied with $S=0.15$ for both states.
The data are taken from
refs.~\protect\cite{bob94}\protect (squares)
 and \protect\cite{qui88}\protect (dots).
In the insert we compare the PWIA  (dotted line), the HF
 and the optical potential results. In this case the three curves are
 not scaled with a spectroscopic factor.}
\label{fig4}
\end{figure}

\begin{figure}
\centering
\epsfxsize=9.5cm
\epsffile{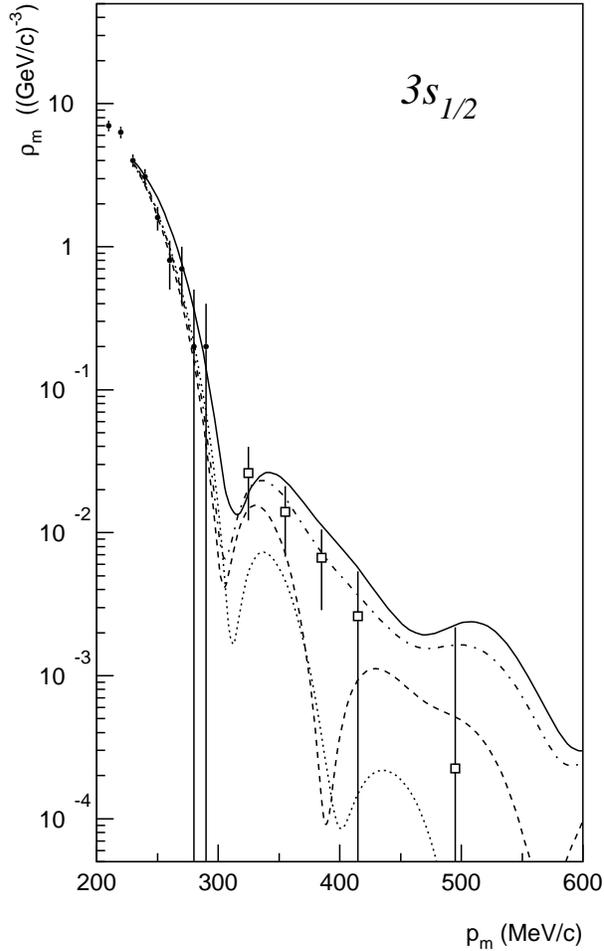}
\caption{Systematic study of the high-momentum components in the
$^{208}$Pb(e,e$'$p) cross section for proton knockout from the $3s1/2$
shell
($\epsilon=487$ MeV, $q=221$ MeV$/c$, $\omega=110$ MeV). 
The data are taken from 
refs.~\protect\cite{bob94}\protect (squares)
 and \protect\cite{qui88}\protect (dots).
The dotted curve represents the results from the HF calculation.
The dashed, dashed-dotted 
and solid line result from, respectively, the RPA,
RPA+MEC+$\Delta$, RPA+MEC+$\Delta$+Coulomb distortion calculation.
All the curves are multiplied with $S=0.15$. }
\label{fig5}
\end{figure}

\begin{figure}
\centering
\epsfxsize=9.5cm
\epsffile{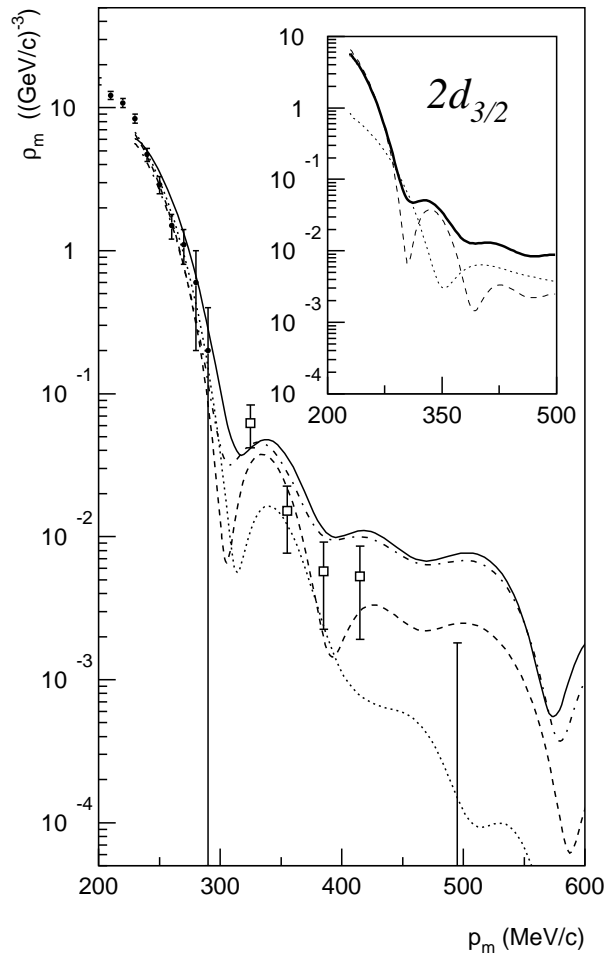}
\caption{Same as Fig.~\protect\ref{fig5}\protect \hspace{.2cm} but for
proton knockout from the $2d3/2$ shell. In the insert 
we plot the one-body  (dashed line) and two-body
 (dotted line) current contribution to the reduced $^{208}$Pb(e,e$'$p)
cross section. The solid
line is the coherent sum of the two curves.
}
\label{fig6}
\end{figure}

In Figs.~\ref{fig4}, \ref{fig5} and \ref{fig6} we plot 
the reduced cross section
for proton knockout from the $3s1/2$ and $2d3/2$ shell in $^{208}$Pb.
The kinematics coincide with the experimental conditions
from ref.~\cite{bob94}.

It goes without saying that  the large
number of protons and neutrons 
makes the $^{208}$Pb(e,e$'$p) 
calculations very involving when accounting for two-body
photoabsorption mechanisms. Moreover, for a nucleus 
containing $82$ protons electron distortion effects 
can no longer be neglected.

Let's first assume a direct knockout mechanism.
In such a picture, the proton is  ejected 
from  the nucleus in a one-step reaction mechanism after
photoabsorption on the one-body nuclear current. The results are
plotted in Fig.~\ref{fig4}. 
The FSI is handled within two
different approaches: the HF approach and the 
optical potential model (OPM).
In both approaches the bound-state wave functions are evaluated with
the HF mean-field
potential. Whereas in the HF approach the continuum single-particle 
states are derived from the
same real HF potential,  in the optical potential calculation  the
scattering potential is a combination of a real and an imaginary 
 potential constructed from the general
parametrization of elastic proton scattering data by
 Schwandt {\em et al.} \cite{sch82}. 
Both models for the FSI yield very similar quantitative results for
the reduced cross secion. 
However, we have to
keep in mind that a considerably smaller reduction  factor
($S(3s1/2)=0.15$, $S(2d3/2)=0.15$) is
adopted for the HF result compared to the spectroscopic factor
($S(3s1/2)=0.51$, $S(2d3/2)=0.54$)
used for the optical potential calculation.
This large difference 
can best be illustrated with  the insert in Fig.~\ref{fig4}. 
The optical potential calculation causes a quenching of the
PWIA (e,e$'$p) strength at the maxima of the reduced cross section. 
In addition, one observes
considerably less pronounced minima. 
It is obvious that due to the lack
of an absorptive imaginary potential, the HF approach 
is not able to reproduce
this strong quenching effect 
but predicts a similar
shape of the reduced cross section as derived with the OPM.
In the HF approach, the transparency, i.e. the
probability that the struck proton emerges from the nucleus without a
collision with the other nucleons in the nucleus, is considered to be one.
However, experimental predictions for the transparency of a nucleus
extracted from (e,e$'$p) experiments vary from $0.8$ to $0.4$ for
target nuclei in the range of $A=12$ to $A=181$ \cite{gar92,gee89}.
These numbers can be reproduced within the model of Pandharipande and
Pieper \cite{pan92} adopting the correlated Glauber approximation and
accounting for a density dependency of the nucleon-nucleon interaction.
 This means that 
for heavy target nuclei the
absorptive part of the optical potential causes a strong 
 reduction of
the cross section by a factor $2.5$, in this way explaining the small
spectroscopic factor extracted within our HF approach.
On the other hand, for light nuclei, the OPM and the HF approach are
expected to produce spectroscopic factors that do not differ by more
than 20\%.
The $^{16}$O(e,e$'$p) analysis reported in
ref.~\cite{vsl94} seems to confirm this observation.

An important advantage of the HF approach is the fact that
 both bound and continuum single-particle states are
derived from the same real mean-field potential, respecting in this
way orthogonality between the continuum  and bound states.
It is well-known that any deviation of this orthogonality 
requirement might cause substantial
spurious contributions to enter the (e,e$'$p) cross sections and that
this problem is predominantly manifest at high
$p_m$ \cite{bof84}. 
In order to avoid these complications, 
we  consider the HF-RPA approach as a good starting point to study the
relative importance of higher-order mechanisms in the quasielastic
(e,e$'$p) reaction.

From Fig.~\ref{fig4} it is clear that the adopted direct knockout
picture underestimates   the measured (e,e$'$p) strength at
high missing momenta ($p_m > 300$ MeV$/c$). 
In conformity with the results for the
$^{16}$O(e,e$'$p) reaction, we expect that
multi-step processes of the RPA-type, photoabsorption on two-body
nuclear currents and Coulomb distortion (CD) effects will
modify the reduced cross sections.
The following  discussion is meant to find out the {\em relative}
importance of these higher-order effects
 to the (e,e$'$p) cross section.


In going from the HF approach with only one-body components in the
nuclear current to a more complete calculation that includes long-range
correlations of the RPA-type, two-body nuclear currents and electron
distortion effects (Figs.~\ref{fig5} and \ref{fig6}),
 an enhancement of strength, especially in
the high-momentum tail of the cross section, is observed. 
The $^{208}$Pb(e,e$'$p)
experiment was done in the high-energy tail of the
quasielastic
peak. Therefore,
 it is not surprising that processes beyond the direct
knockout picture come into play.   
All the curves in the Figs.~\ref{fig5} and \ref{fig6}
are multiplied with one and the same scaling factor ($S=0.15$). 

The role of RPA correlations is
mainly manifest 
at the
high-momentum side of the reduced cross section.
A possible explanation is the following. Multi-step processes tend to
redistribute the strength over a wider missing momentum range and,
consequently, shift some strength from the lower to the higher $p_m$
region \cite{kel94}. Indeed, after a number of rescattering processes
the detected kinetic energy of the escaping nucleon does no longer
uniquely determine the momentum of the nucleon on which the initial
photoabsorption took place.
Combining both considerations the reduced cross section does not 
 scale as a function of the missing momentum when RPA correlations
are accounted for and the (e,e$'$p) strength is smeared 
out towards higher
missing momenta.

In a similar way one can explain the substantial contribution of
two-body nuclear currents to the reduced cross
section.
Electro-induced one-proton knockout after
photoabsorption on a two-body nuclear current also generates 
a considerable
amount of strength at the high missing momentum side of the reduced
cross section. As 
demonstrated in the insert of Fig.~\ref{fig6}, 
a coherent sum of the one-body and two-body nuclear
current  contributions is required. 
The pure two-body contribution shows a much
smoother behaviour as a function of $p_m$ compared to the one-body
part.
This can be easily explained from the fact that for the two-body
absorption mechanism the missing momentum $p_m$ no longer serves as a
scaling variable.
 Moreover, whereas for the low missing momentum side the strength
generated by the two-body mesonic currents is at least 
an order of magnitude
smaller than the pure nucleonic contribution (=one-body nuclear
current), for the highest missing momenta 
the two-body strength {\em can}
overshoot the one-body contribution. 
Consequently, the reduced cross
section at the low missing momentum side can be mainly attributed to
proton knockout after photoabsorption on a one-body nuclear current.
To the contrary, two-body current contributions and
interference effects between the one- and two-body
part in the nuclear current 
come into play for the 
high missing momentum side of the reduced cross section.

In the paper by Bobeldijk {\em et al.} \cite{bob94}
 the effect of mesonic currents
is estimated at 10\% of the one-body contribution. This number is
based on a calculation for $^{40}$Ca \cite{bof90}. However, 
as demonstrated for the
target nuclei
 $^{16}$O (Fig.~\ref{fig2}) and  $^{208}$Pb
(Figs.~\ref{fig5} and \ref{fig6}), the
relative contribution of two-body nuclear currents to the reduced
cross section exhibits a clear
mass-dependence. At high missing momenta, 
mesonic currents seem to be relatively more
important for the $^{208}$Pb(e,e$'$p) 
compared to the  $^{16}$O(e,e$'$p) cross sections when
considering comparable
kinematical conditions. This observation is not too surprising.
In contrast with the particle-hole matrixelement 
for the one-body current,
the matrixelements for the two-body currents 
 involve a
summation over the different occupied states in the target nucleus
\cite{vsl94}.  In this way, a clear mass-dependence is introduced in
the two-body current contribution.

Considering the (e,e$'$p) reaction from a heavy nucleus like 
$^{208}$Pb, Coulomb distortion effects
cannot be discarded.
It is clearly seen from
Figs.~\ref{fig5} and \ref{fig6} that
electron distortion effects modify the shape of the 
reduced cross
sections for proton knockout from the two considered shells.
The size of the corrections related to Coulomb distortion depends on
the quantum numbers of the shell from which the proton is ejected.  
We notice that
the $3s1/2$ reduced cross section is more affected by electron
distortion effects than the $2d3/2$ result.
For the two single-particle states, 
the Coulomb distortions induce strength at
high missing momenta and fill up the minima of the reduced cross sections.

We may conclude that all three higher-order effects taken into
consideration in our study, all 
improve on the
agreement with the data as they account for part of 
the missing (e,e$'$p)
strength  at high missing momenta.

In a recent paper by Bobeldijk {\em et al.} \cite{bob94} the
${}^{208}$Pb (e,e$'$p) data at high missing momenta were 
analyzed within the
framework of the Coulomb Distorted Wave Impulse Approximation (CDWIA)
 approach of the Pavia group \cite{bof93}. 
The bound-state Wood-Saxon wave functions adopted in this model
were modified with various types
of correlation functions. 
These correlation functions account in a semi-phenomenological way for
short-range and long-range nucleon-nucleon correlations which are not
implemented in the mean-field single-particle wave functions.
In conformity with the predictions of M\"{u}ther {\em et al.}
\cite{mut94}, Bobeldijk
{\em et al.}
arrived at a small effect of the SRC at the low missing energies
probed in the experiment.
They attributed the mismatch between the CDWIA (no correlation
functions included) approach and the
data to long-range effects in the nuclear wave function.
These results partially agree with our results.
Indeed, as can be learned from our calculations 
long-range correlations (of
the RPA-type)  considerably contribute to the cross section at
high missing momenta.
In our model, however, the calculated mesonic contribution 
to the reduced cross section   is found to be 
considerably larger than the estimation reported 
in the paper by Bobeldijk {\em et al.}.

Recently, also a fully relativistic analysis of the $^{208}$Pb(e,e$'$p)
results at high $p_m$ has become available \cite{udias}.  In comparison
with the non-relativistic approaches the relativistic models reduce
the quasielastic (e,e$'$p) cross section at low missing momenta
\cite{udias2,udias}.  At high missing momenta the opposite behaviour
is noticed.  The degree to which these effects occur is, however,
strongly dependent on the adopted choice for the relativistic
off-shell nuclear current operator.  The relativistic calculations of
ref.~\cite{udias} made further clear that the predicted sensitivity to
the choice of the current operator is more pronounced at high than at
low $p_m$.  In that sense, the conclusions are similar to those drawn
within our non-relativistic treatment.  The off-shell behaviour of the
relativistic nuclear current operator could indeed be considered as an
effective way of accounting for the many-body correlations (like
ground-state correlations, meson-exchange currents, ...) in the
(e,e$'$p) reaction process.  In that respect it is worth mentioning that
the qualitative effect of the RPA ground-state correlations on the
(e,e$'$p) cross sections of Figs.~5 and 6 is similar to the effect being
ascribed to relativity in ref.~\cite{udias} : a (modest) decrease of
the strength at low $p_m$ and a considerable increase at high $p_m$.

\section{Conclusions}

In this paper we have reported on the 
(e,e$'$p) reaction   at high
missing momenta and low missing energies. 
We confronted our $^{208}$Pb results with data taken at NIKHEF. 
This experiment was primarily meant to investigate the 
short-range
nucleon-nucleon correlations in the nucleus
 as they are known to show up at high
momenta in the nucleon momentum distribution.
Starting from the
direct knockout picture for one-proton knockout reactions, our model
implements higher-order effects such as
correlations of the RPA-type,
 photoabsorption on two-body one-pion
exchange currents and electron distortion effects in a systematic and
consistent way.
It has been shown that all these effects strongly modify the
reduced cross section at high missing momenta.
Notwithstanding the fact that the direct knockout picture is
sufficient to reproduce the
low missing momentum side of the reduced cross section,
these higher-order effects need to be carefully examined
before any conclusions on the role of nucleon-nucleon correlations
for the high-momentum components in the nucleon
momentum distribution can be drawn from (e,e$'$p) data.

{\bf Acknowledgement}

The authors are grateful to K. Heyde for fruitful discussions and
suggestions. 
This work has been supported by the Inter-University
Institute for Nuclear Sciences (IIKW) and the National Fund for
Scientific Research (NFWO).

\end{document}